\documentclass[preprint,endfloats*,a4paper,byrevtex,aps]{revtex4}

\usepackage{graphicx}
\newcommand{\etal}{\textit{et al.}}

\begin{document}

\title[Nanofriction on corrugated samples - preprint]{Quantitative Nanofriction
Characterization\\ of Corrugated Surfaces by Atomic Force
Microscopy}\altaffiliation{Submitted to Journal of Applied Physics, Feb. 2003}
\date{\today}
\author{A. Podest{\`a}}
\author{G. Fantoni} \altaffiliation[Present address: ]{Agusta, Lab.
Tecnol. Materiali - Controlli Non Distruttivi -
via G. Agusta, 520 Cascina Costa di Samarate (VA), Italy}
\author{P. Milani}\email{pmilani@mi.infn.it}
\affiliation{INFM - Dipartimento di Fisica, Universit{\`a} degli Studi di Milano,\\
Via Celoria 16, 20133 Milano, Italy}

\begin{abstract}
Atomic Force Microscopy (AFM) is a suitable tool to perform tribological
characterization of materials down to the nanometer scale. An important aspect in
nanofriction measurements of corrugated samples is the local tilt of the surface,
which affects the lateral force maps acquired with the AFM. This is one of the most
important problems of state-of-the-art nanotribology, making difficult a reliable and
quantitative characterization of real corrugated surfaces. A correction of topographic
spurious contributions to lateral force maps is thus needed for corrugated samples. In
this paper we present a general approach to the topographic correction of AFM
lateral force maps and we apply it in the case of multi-asperity adhesive contact. We
describe a complete protocol for the quantitative characterization of the frictional
properties of corrugated systems in the presence of surface adhesion using the AFM.
\end{abstract}

\maketitle


\section{Introduction}

The study of tribological properties at sub-micrometer scales has been stimulated by
the reduction of dimensions of mechanical devices. Micro Electro-Mechanical Systems
(MEMS), Nano Electro-Mechanical Systems (NEMS), or hard disk drives
require an accurate control of friction, wear, and adhesion at the
nanometer scale~\cite{mrsbull01,bhu99b,bhu01}. From a fundamental point of view the investigation
of the influence of the chemico-physical condition of the surface at the nanoscale on
the tribological behavior can provide important elements for the understanding of friction, wear, and
lubrication phenomena.

Atomic Force Microscopy (AFM) is one of the most powerful techniques for the investigation
of tribology and in particular of nano-friction~\cite{carp97,ded00,gne01}. An atomic force
microscope can simultaneously acquire topographic maps of surfaces with
nanometric resolution, and friction maps operating in the so-called Friction-Force
mode. Friction Force Microscopy (FFM) is possible thanks to the simultaneous
acquisition of the vertical deflections of the beam supporting the AFM tip, related
to changes in the topographic relief, and the lateral deflections, proportional to
the friction force between the tip and the sample surface~\cite{mey90,bhu99b}.

In order to perform quantitative friction measurements it is
necessary to control and measure accurately the forces acting on the AFM tip.
Parameters such as roughness, granularity, power spectrum of the surface play an
important role. The effects of surface topography on nano-friction measurements have
been studied, although a general theory is still lacking~\cite{graf93,lab94b,lab94,koi97,sun00}. In many
cases, attention has been concentrated on flat crystalline surfaces under
ultra-high-vacuum (UHV), where the influence of surface topography is
negligible~\cite{car96,schw95,ena98,gne01b}.

On the other hand, real surfaces in every-day conditions necessitate to take into
account the topographic contribution to lateral force maps in order to achieve
quantitative tribological information. Actually, in the case of corrugated samples,
that is of locally tilted surfaces, the measured forces in the directions parallel and
perpendicular to the AFM reference plane do not necessarily coincide with the forces
acting parallel and perpendicularly to the sample surface, which actually define the friction
coefficient and the friction vs. load characteristics of the interface under
investigation~\cite{graf93,lab94b,lab94,koi97,sun00}. These effects related to surface morphology depend on the ratio
between the dimension of the sliding probe and that of the surface asperities. In
this sense friction is a scale-dependent phenomenon, and a tool such the AFM, which
is able to provide morphological information on a wide range of scales, is very
attractive for such studies.

In this paper we consider the role of the topographic correction in a general way and inquiry
whether it is possible to follow a model-independent approach, providing the
friction vs. load characteristics of the system under investigation without the need
of postulating any particular contact-friction model. We discuss the experimental
limitations, which make the model-independent approach unreliable and we solve the
problem of the topographic correction in the particular case of the adhesive
multi-asperity contact, which is of common occurrence in many experimental setups.
To this purpose, we introduce a modified version of the Amonton's law for
friction~\cite{bow50} (linear dependence of friction on the load), which should better
apply to the case of low loads and few asperities in contact. We present our
topographic correction procedure in the framework of a complete quantitative
statistical protocol based on the AFM for the characterization of frictional
properties of materials at sub-micron scale. We discuss the limit of validity of
this model and highlight the underlying assumptions.

The article is organized as follows: in sections~\ref{section:topo_correction}
and~\ref{section:instr_offset} we present the theory of the topographic correction
of lateral force maps acquired with a standard beam deflection AFM; in
section~\ref{section:adhesion} we discuss the multi-asperity adhesive contact regime
typical of FFM experiments and present a friction model which should apply to FFM
experiments in the presence of humidity; in section~\ref{section:topo_corr_Amontons}
we include this model in the topographic correction theory and derive the basic
equations for interpreting the experimental data; in section
\ref{section:protocol_exp} we describe the experimental setup and the
characterization protocol; in section~\ref{section:tribo_exp} we discuss the
validity of the topographic correction protocol via its application to a corrugated
Polytetrafluoroethylene surface. Conclusions are reported
in section~\ref{section:conclusion}.

\section{\label{section:topo_correction}FFM on corrugated samples: the topographic
correction}

\begin{figure}[h]
 \centering
 \includegraphics[scale=1]{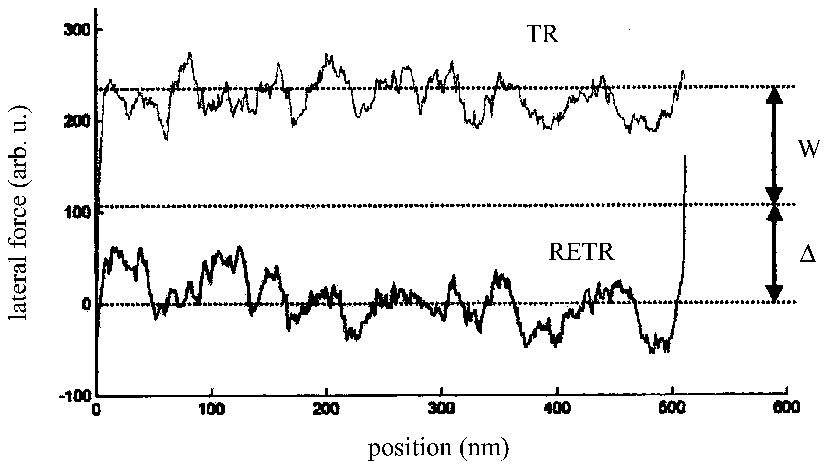}
 \caption{\label{fig:jap1}}
\end{figure}

In a typical FFM experiment the cantilever is scanned across the surface in a
direction orthogonal to the long cantilever axis. Using a segmented photodiode, with
two upper and two bottom quadrants, it is possible to measure accurately not only
the vertical cantilever displacements, but also the cantilever lateral torsion. The
vertical deflection of the cantilever is proportional to the applied load, while
fluctuations around the average position are due to corrugations of the sample
surface~\cite{bhu99b}. These vertical fluctuations represent the input of a feedback loop, keeping the cantilever deflection, that is the applied load, constant. Recording the relative
displacements of the tip-sample assembly provides the AFM topographic map. The
torsion of the cantilever is in turn proportional to the friction force between the
AFM tip and the sample surface. The acquisition of lateral cantilever deflections
(the so-called lateral force signals) provides the friction map. Both topographic
and friction maps can be recorded simultaneously. Fig.~\ref{fig:jap1} shows
a typical friction loop, i.e. the lateral force signals $TR$ and $RETR$ (in
arbitrary units) acquired in the two opposite directions during a single scan (these
directions are usually called Trace and Retrace). The typical loop shape is due to the
fact that the lateral signal changes sign when the motion of the tip is reversed. The
ramps at the beginning of the Trace and Retrace curves are due to a \textit{pivot}
effect during the reversal of the tip motion.

\begin{figure}[h]
 \centering
 \includegraphics[scale=1]{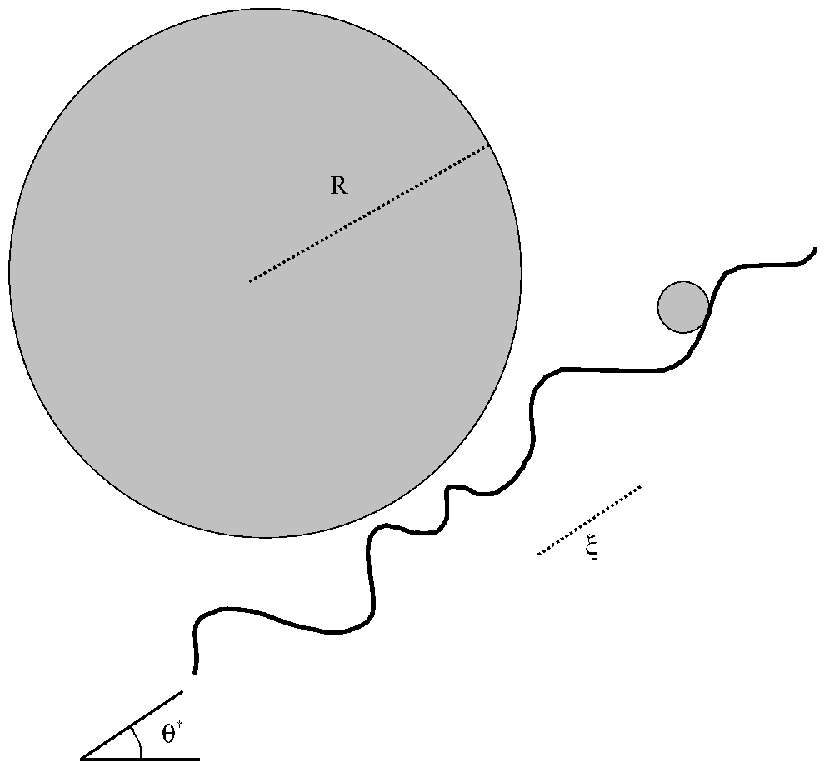}
 \caption{\label{fig:jap2}}
\end{figure}

In order to achieve a quantitative understanding of the friction properties of a
surface one needs to consider closely the contact regime occurring during the
experiment. Contact regimes in FFM and standard macroscopic friction tests can be very
different~\cite{bhushan01,bhu01b}. In typical FFM measurements, the radius of curvature of the AFM tip can be comparable or even smaller to the size of
morphological surface features (few nanometers). If $\xi$ is a measure of the
average extension of morphological features of the surface (the correlation length
of the surface can be used~\cite{bar95}) and $R$ is the AFM tip radius, the
aforementioned situation is characterized by the condition $\xi/R\geq1$
(Fig.~\ref{fig:jap2}, right). In this case a single-asperity contact is to
be expected~\cite{ada00}. The limiting case is represented by a sharp spherical or
parabolic tip (as those prepared via electron beam deposition techniques~\cite{schw97})
sliding on an atomically smooth surface (a situation which can be well realized in
UHV-FFM apparatus).
In macroscopic friction experiments a slider, typically a ball or a disk, is scanned
across a surface with a geometric or apparent contact area as large as few
$cm^{2}$~\cite{bhu01b}.
As long as the dimension of the slider is large compared to $\xi$
($\xi/R\leq1$, Fig.~\ref{fig:jap2}, left), a multi-asperity contact is to be
expected~\cite{ada00}.

The friction coefficient $\mu$ is defined as the ratio $f/N$, where $f$ is the
opposing friction force acting on the slider in the direction perpendicular to the
local surface normal and $N$ is the load applied along the surface normal. The
presence of a global tilt of the surface, that is a tilt on a scale comparable or
larger than that of the probe, can affect the measurement of the friction
coefficient. This is because in typical friction apparatus the forces tangential and
perpendicular to the ideally flat laboratory reference plane are measured, say $T$
and $L$. If the surface is tilted, $T$ and $L$ are no longer parallel to $f$ and $N$,
accordingly. A topographic correction to the measured quantities $T$ and $L$ is thus
necessary in order to get the values of $f$ and $N$ and from these the value of the
friction coefficient~\cite{graf93,lab94b,lab94,koi97,sun00}. In many experimental situations,
typically in macroscopic friction tests, the sample surface is supposed to be flat on
a length scale large compared to the slider dimension. In this case the measured
forces corresponds to the local normal and tangential components, and the above
definition of the friction coefficient readily applies. In FFM experiments on
corrugated surfaces however the probe is likely to be scanned on a locally tilted
surface, as shown in Fig.~\ref{fig:jap2}. The topographic correction is thus
an essential part of any protocol devoted to the quantitative characterization of
frictional properties of surfaces.

A good parameter characterizing the flatness of the sample surface is the local
slope $\tan(\theta)_{R}$. The pedex $R$ means that the slope is calculated with
resolution equal to the slider dimension. For the surface to be considered locally
tilted, it is necessary that the tilt extends on a length bigger than the slider
dimension (i.e. $\tan(\theta)_R\neq0$), as shown in Fig.~\ref{fig:jap2},
where the local tilt of the surface, $\theta^*$, is calculated with resolution equal
to the radius of the bigger probe. The map of local slopes is readily obtained from
the topographic map. Because of the possibility of acquiring simultaneously the
topographic and the lateral force maps, FFM is the technique of choice to perform
accurate friction measurements at those small scales where the topographic correction is required.

\begin{figure}[h]
 \centering
 \includegraphics{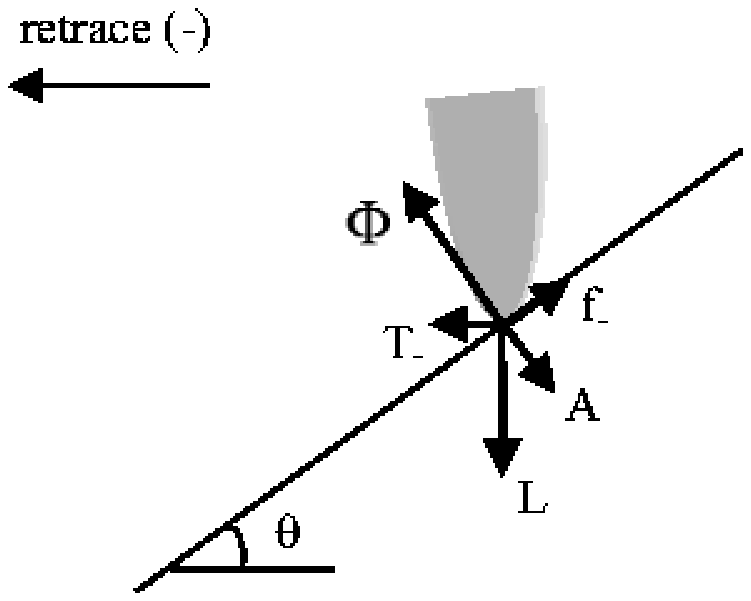}
 \caption{\label{fig:jap3}}
\end{figure}

In order to consider the topographic correction, we show in Fig.~\ref{fig:jap3} the
forces acting on the AFM tip sliding over a sloped surface. The negative sign indicates the
retrace direction. Notice that $\theta\equiv \theta_{R}$ is the local surface tilt.
The tip is forced against the object by two forces exerted by the cantilever: one is
the applied load $\mathbf{L}$\footnote{Notice that $\mathbf{L}$ is the
\textit{externally} applied load, as measured from the deflection of the cantilever,
controlled by the deflection setpoint.}, directed downward, the other is the tangential force $\mathbf{T}$
originating in the torsion of the cantilever, in the direction of motion. Three
other forces act on the tip: the friction force $\mathbf{f}$, perpendicular to the
local normal and always opposing the motion, the local surface adhesion
$\mathbf{A}$, directed as the local normal, and a normal reaction $\mathbf{\Phi}$,
equal in magnitude but opposite in direction to the \textit{total} applied load
$\mathbf{N}$. It should be noticed that while the applied load $\mathbf{L}$, the
adhesion force $\mathbf{A}$ and the surface reaction $\mathbf{\Phi}$ have always the
same direction, the tangential and friction forces $\mathbf{T}$ and $\mathbf{f}$
invert their direction as the motion is reversed, from Trace to Retrace.

Projecting the equilibrium condition
$\mathbf{T}+\mathbf{f}+\mathbf{L}+\mathbf{\Phi}+\mathbf{A}=\mathbf{0}$ for Trace (+)
and Retrace (-) along the local normal and tangential axes provides a system of four
equations, which represents the topographic correction, providing $f$ and $N$ as a
function of $T$, $L$, $A$ and $\theta$:
\begin{eqnarray}
f_{+} & = & T_{+}\cos(\theta)-L\sin(\theta)\nonumber \\
N_{+} & = & L\cos(\theta)+T_{+}\sin(\theta)+A\nonumber \\
\vspace{.2cm}
f_{-} & = & T_{-}\cos(\theta)+L\sin(\theta)\nonumber \\
N_{-} & = & L\cos(\theta)-T_{-}\sin(\theta)+A \label{eq:topo_correction_TR-RETR}
\end{eqnarray}
Given a number of $(f,N)$ pairs, the model-independent friction vs. load curve of
the interface considered is obtained. Depending on the mechanics of the contact
(single or multi-asperity), both a power law and a linear behavior can be
observed~\cite{ada00}.
Interesting physical parameters such as the friction coefficient, the surface
energy, the true contact area etc. can in principle be extracted fitting the
experimental $f$ vs. $N$ curve.

In the above equations $T$ is the force acting parallel to the AFM reference plane.
In fact, what is measured are the true lateral force signals, $T_{\pm}$, plus an
instrumental offset $T_{instr}$ ($TR$ and $RETR$, which form the friction loop shown
in Fig.~\ref{fig:jap1}). Assuming the instrumental offset to be
independent on the direction of motion, we have:
\begin{eqnarray}
TR & = & T_{+}+T_{instr}\nonumber\\
RETR & = & -T_{-}+T_{instr}\label{tr_retr}
\end{eqnarray}

We can now reconsider more closely the friction loop in Fig.~\ref{fig:jap1}
in order to express the equations of the topographic correction in terms of
measurable quantities. A vertical offset $\Delta$ is present in the friction loop,
causing the loop having a non-zero average. The offset $\Delta$ of the friction loop
is calculated as:
\begin{equation}
\Delta=1/2\cdot(TR+RETR) \label{eq:Delta}
\end{equation}
Also indicated in Fig.~\ref{fig:jap1} is the average half width of the
loop, $W$:
\begin{equation}
W=1/2\cdot(TR-RETR) \label{eq:W}
\end{equation}
Substituting Eqs.~\ref{tr_retr} into Eqs.~\ref{eq:Delta} and~\ref{eq:W} we obtain:
\begin{eqnarray}
W & = & \frac{T_{+}+T_{-}}{2}\nonumber\\
\Delta & = & T_{instr}+\frac{T_{+}-T_{-}}{2}\label{eq:W_Delta}
\end{eqnarray}
We see from Eqs.~\ref{eq:W_Delta} that even if we were able to subtract
point-by-point the instrumental offset from the friction loop, still the loop
average would be non-zero, but equal to $(T_{+}+T_{-})/2$. This residual offset is
intrinsically due to the presence of a local non-zero slope, causing the local
effective load $N$, and consequently the local friction $f$, and the lateral signals
$T_{+}$ and $T_{-}$ to be different in Trace and Retrace (see
Eqs.~\ref{eq:topo_correction_TR-RETR}). Moreover, it turns out that
$W$ is an approximation of both the Trace and Retrace lateral force signals $T_{+}$ and
$T_{-}$, i.e. an average lateral force. While the instrumental offset cancels out in Eq.~\ref{eq:W}
after the subtraction\footnote{The instrumental offset cancels out
under the hypothesis that it does not change in Trace and Retrace.}, there are
spurious contributions to lateral force which do not cancel, having the same sign (but
generally different magnitudes) in Trace and Retrace. These contributions are due to
the \textit{ploughing} and \textit{collision} mechanisms reported for instance in Ref.~\onlinecite{sun00}.
These spurious contributions are related to sudden changes in surface slope, but they
can not be accounted for as we did for the other topographic contributions discussed
above in that there is not any analytic relation to parameters such as slope,
applied load and scanning velocity. However, provided the scanning velocity is not
very high and the feedback loop of the AFM is properly adjusted, these spurious
contributions to lateral force maps can be smoothed out if a large number of points
in the map are acquired, in that they represent only a small fraction of the lateral
force map.

Finally from Eqs.~\ref{eq:W_Delta} we obtain:
\begin{eqnarray}
T_+ & = & W+\Delta-T_{instr}\nonumber\\
T_- & = & W-\Delta+T_{instr}\label{T+_T-}
\end{eqnarray}

Substituting Eqs.~\ref{T+_T-} into Eqs.~\ref{eq:topo_correction_TR-RETR}, the
model-independent friction characteristics $f$ vs. $N$ in terms of the measurable
quantities $W$ and $D$ and the instrumental offset $T_{instr}$ is obtained. In the
following section we will discuss the properties of the instrumental offset and
highlight the reasons why its measurement is not usually reliable.

\section{\label{section:instr_offset}Instrumental offset}

The instrumental offset $T_{instr}$ is determined by thermal drifts of the cantilever,
electronic noise, laser intensity fluctuations and laser interference, changes in
the environmental conditions (temperature, relative humidity) and cross-talking
effects, causing a spurious lateral signal even in the absence of any true lateral
force. The instrumental offset can vary from scan to scan, and also during the same scan,
and it is in general load-dependent. Cross-talking represents the most important
contribution to the instrumental offset~\cite{rua94,ogle96}. It is caused by a bad
alignment of the laser beam with respect to the photodiode array. If the plane
containing the beam is not perpendicular to the plane of the photodiodes, vertical
displacements of the cantilever will cause a spurious lateral signal, even in the
absence of any applied load and/or friction force. The alignment of the laser with respect to the
cantilever and photodetector is thus a critical step in any FFM experiments. Some
authors report about correction of the cross-talking effect via electronic
compensation~\cite{ogle96}, however this imply the risk of manipulating the true signal and it is
in general not simple to perform, in that it requires an accurate measurement of the
spurious lateral force signal versus the applied load.

\begin{figure}[h]
 \centering
 \includegraphics[scale=1]{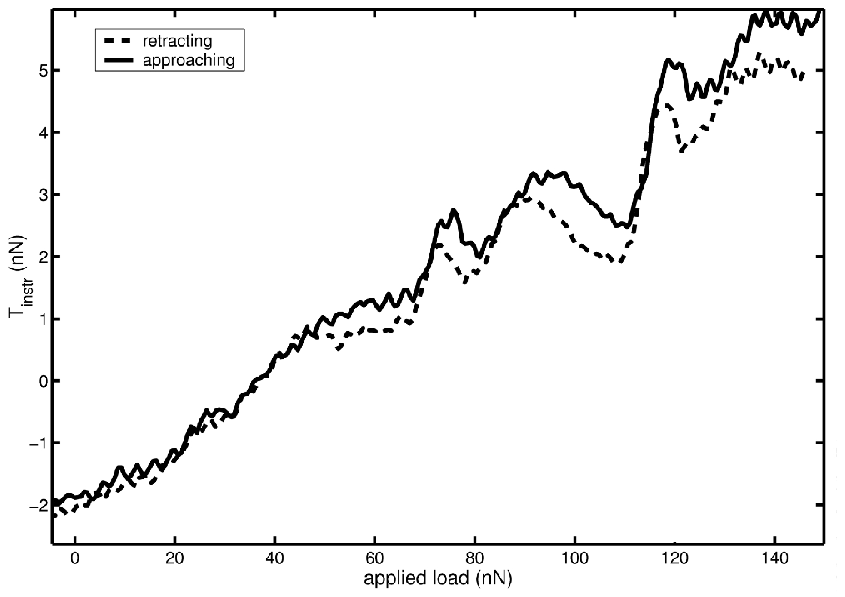}
 \caption{\label{fig:jap4}}
\end{figure}

One way to characterize the instrumental offset is to record the lateral force signals
during approaching-retracting cycles of the tip on the sample.
Because the cantilever does not move laterally during approaching-retracting cycles,
any lateral deflection measured by the photodiodes can be attributed to
cross-talking effects. The force vs. distance curve and the lateral force curve are
in one-by-one correspondence: the first provide the applied loads, while the second
provides the spurious lateral force offset corresponding to each applied load. In
Fig.~\ref{fig:jap4} the $T_{instr}$ vs. applied load curve acquired with the
method described above is shown.
The signals acquired in the approaching and retracting part of the
cycle are different. Moreover, the spurious lateral force is not negligible, compared with
typical frictional forces in FFM experiment (few nN). The presence of strong
laser interference is visible.

It is definitely clear in Fig.~\ref{fig:jap4} that the instrumental offset does depend in general
on the applied load. This method is useful to qualitatively characterize the
instrumental offset and its load dependence. However, it is not reliable. Even if one
takes the average over many approaching-retracting cycles, the offset changes quickly,
and the curve acquired before the FFM measurement is usually different from the one
acquired after. Moreover, the bending and twisting of the cantilever during the
approaching-retracting cycles does not mimic necessarily those occuring during the
Trace and Retrace scans for any given applied load, if not qualitatively. In
principle, a more reliable method to characterize quantitatively the instrumental
offset could be to measure the spurious lateral deflections during free oscillations
of the cantilever, at different oscillation amplitudes, which would correspond to different
applied loads in contact mode. The drawbacks of this method, not implemented here,
are that only relatively small amplitudes can be set (i.e. low loads), and that the
tip is not in contact with the sample, that is, there could be contact-induced
contributions to the offset, which are not taken into account when measuring the free
oscillations of the cantilevers.

The topographic correction (Eqs.~\ref{eq:topo_correction_TR-RETR}
and Eqs.~\ref{T+_T-}) based on the direct measurement of the instrumental offset
is hampered by the difficulty in measuring reliably and accurately this essential
parameter. A possible way to circumvent this problem is to fix, a priori, the friction
regime (in fact, only the multi-asperity case can be worked out analytically, because
of the simple linear dependence of friction on load). Apart from difficulties in the
calculations, even for the simplest models, it must be pointed out that this strategy
is not fully satisfactory, because it assumes the friction regime, while it would be
better to infer it from experimental data. However, the validity of the model adopted
can be confirmed, a posteriori, from the shape of the lateral force vs. load curves, and
the choice of the model can be supported by reasonable assumptions about the material to
be investigated and the size and shape of the AFM tip.

\section{\label{section:adhesion}Adhesion and the modified Amonton's law}

In this section we address the problem of the adhesive contact between two surfaces,
in order to define and possibly justify a reasonable friction model to be used in
the equations of the topographic correction developed in the previous paragraphs.

A fundamental relation is the proportionality between
the friction force $f$ and the contact area $a$ between two bodies~\cite{bow50}:
\begin{equation}
f=\tau a \label{eq:fpropA}
\end{equation}
$\tau$ being the shear modulus, i.e. the friction force per unit area.
Eq.~\ref{eq:fpropA} has been verified experimentally both directly (with the Surface
Force Apparatus~\cite{isr92} or friction-current measurements~\cite{lan97}) for
single-asperity contacts and indirectly, through the validity of the contact
mechanics models~\cite{ena98}.
The shear modulus $\tau$ depends in general on the number and nature of bonds forming
at the interface, while the contact area depends on the mechanical properties of the
interface, that is on the bulk properties of tip and sample materials.

Contact mechanics aims to provide for every contact regime the functional relation
$a(N)$ between the contact area $a$ and the load $N$. Once the relation $a(N)$ is
established, the friction vs. load characteristics $f(N)$ for a particular system is
given by Eq.~\ref{eq:fpropA}. Classical contact mechanics for continuous elastic
bodies was proved to be applicable to nanometer-sized contacts, as the ones
occurring in FFM~\cite{car96,schw97,fra97,lan97}.

FFM offers a wide spectrum of contact regimes, depending on both interface
properties (tip radius, surface corrugation, as well as elastic properties of tip
and sample), and environmental properties, such as relative humidity. The load range
is also important in determining the contact regime. During FFM experiments on flat
surfaces using well characterized spherical or parabolic tips a single-asperity contact is likely to take place~\cite{schw97}. When
corrugated samples are investigated in ambient conditions, in the presence of humidity and
contaminants, a multi-asperity contact is even possible, altough the smoothing effect
of contaminants of tip and sample nano-corrugations, leading to a transition from a
multi- to a single-asperity contact, has also been observed~\cite{put95}. The AFM
allows indeed to investigate different contact regimes, provided a good
characterization of the tip and sample morphology is given and the possibility of
imaging under controlled environmental conditions.  

In the case of non-adhering elastic and plastic multi-asperity contact, a linear
dependence of friction on load $L$ is expected, the so-called Amonton's
law~\cite{bow50}:
\begin{equation}
f=\mu L \label{eq:amontons}
\end{equation}
Given a load in the range [0-100] nN and a typical contact area from a few
tens to a few thousands of nm$^2$, a pressure in the MPa-GPa range is applied to
the sample. This range is comparable to those of macroscopic friction tests, where
loads are in the Newton range and contact areas can be up to few cm$^2$~\cite{bhu01b}. However, in
FFM, load can be comparable to adhesive forces (up to hundreds of nN). This implies
that adhesion plays a fundamental role in every FFM experiment in ambient condition,
and it is expected to strongly influence the friction behavior.

A considerable effort has been made in order to include adhesion in the models of multi-asperity
elastic contact~\cite{ful75,bus76,ken86,per01,cho01}. A definite model of roughness-dependent adhesion is
still lacking. The available studies on adhesive multi-asperity contacts do not
provide the relation between the area of contact and the load, which would provide
in turn the actual friction law through Eq.~\ref{eq:fpropA} in the case of
multi-asperity adhesive contacts. If adhesion is considered explicitly in Amonton's
equations, setting $N=L+A$, where $L$ is the external applied load and $A$ is the
adhesion force, Eq.~\ref{eq:amontons} becomes:
\begin{equation}
f=\mu N=\mu (L+A) \cong \text{const}+\mu L \label{eq:coulomb}
\end{equation}
This form of the Amonton's law is also known as the Coulomb equation~\cite{bow50}.
At zero \textit{external} applied load ($L=0$), Eq.~\ref{eq:coulomb} predicts a
non-zero friction force, as in the case of both the Johnson-Kendall-Roberts
(JKR)~\cite{joh71} and Derjaguin-Muller-Toporov (DMT)~\cite{der71} models for the
single-asperity adhesive contact. Unless the JKR model, however, the Coulomb law does
not account for any finite frictional force in the limit of zero-total applied load,
i.e. at pull-off. Another limitation of the Coulomb law is that adhesion $A$ is
considered constant, while it is reasonable that it varies with load, expecially in
the typical FFM conditions. 

The contact mechanics of an FFM represents an intermediate case between the
macroscopic multi-asperity case and the UHV single-asperity case, in that the
contacting asperities, especially in the low-load regime, can be really a few. The
SPM case lies in the low load, few asperities range where the asymptotic predictions
of the different existing models fail. This point makes the study of the
multi-asperity adhesive regime in FFM experiments very difficult. Without a complete
contact theory, one may try to use reasonable models justifying them a posteriori
with the experimental observations and theoretical results in the single-asperity
limit. \label{page:modified_amontons}A possibility is to generalize Eq.~\ref{eq:coulomb} including an offset $c$ in the linear dependence:
\begin{equation}
f^{\ast} \simeq \mu N +c \label{eq:mod_amonton}
\end{equation}
The linear dependence on load accounts for the fact that the contact is not
single-asperity -like, especially when the AFM is operated on corrugated
samples in humid air. The offset $c$ accounts for the zero-total
load friction force typical of many single-asperity contact, a limit not too far
from the actual experimental situation. Moreover, as pointed out by Schwarz \etal,
an offset in the Amonton's law accounts also for the uncertainty in the
determination of the pull-off force~\cite{schw96,schw95}. Whether the adhesive
offset is due to uncertainty in the measurements of adhesion or is intrinsic, this
can be proved a posteriori on the basis of its numerical value (see discussion
below). Eq.~\ref{eq:mod_amonton} can be thought as the low-load limit of a more
complex $a\equiv a(N)$ dependence. The choice of a modified Amonton's law represents an
attempt to generalize the Coulomb law in order to interprete friction data acquired with the
FFM in common experimental situations, where a multi-asperity adhesive contact is
likely to occur, but with only a few asperities in the low load regime.

\section{\label{section:topo_corr_Amontons}The topographic correction for the modified
Amonton's law}

We will consider Eq.~\ref{eq:mod_amonton} in the general framework of the
topographic correction developed in Section~\ref{section:topo_correction}. We remind
that in Eq.~\ref{eq:mod_amonton} $N$ represents the \textit{total} applied load in
the direction perpendicular to the surface, including the contribution of adhesion
$A$. Substituting Eq.~\ref{eq:mod_amonton} in Eqs.~\ref{eq:topo_correction_TR-RETR}
we obtain a new set of equations giving the values of $T_{+}$, $T_{-}$, $N_{+}$,
$N_{-}$ as a function of the measured local slope $tan(\theta)$\footnote{Hereafter
we assume $\tan(\theta)\equiv tan(\theta)_R$, there $R$ is the tip radius (see
section~\ref{section:topo_correction}).} and applied load $L$, and of the unknown
parameters $\mu$ and $c$:
\begin{eqnarray}
T_{+} & = & L\frac{\mu+\tan(\theta)+(c+\mu A)/(L\cos(\theta))}{1-\mu\tan(\theta)}\nonumber\\
N_{+} & = & L\cos(\theta)\frac{1+\tan^{2}
           (\theta)+(c+\mu A)\tan(\theta)/(L\cos(\theta))}{1-\mu\tan(\theta)}+A\nonumber\\
T_{-} & = & L\frac{\mu-\tan(\theta)+(c+\mu A)/(L\cos(\theta))}{1+\mu\tan(\theta)}\nonumber\\
N_{-} & = & L\cos(\theta)\frac{1+\tan^{2}(\theta)-(c+\mu A)\tan(\theta)
           /(L\cos(\theta))}{1+\mu\tan(\theta)}+A \label{TN_linear}
\end{eqnarray}
Substituting Eqs.~\ref{TN_linear} into Eqs.~\ref{eq:W_Delta}, we obtain a set of equations from whom we aim to extract the
values of the friction coefficient $\mu$ and the offset $c$:
\begin{eqnarray}
W      & = & L\mu\frac{1+\tan^{2}(\theta)}{1-{\mu}^{2}\tan^{2}(\theta)}+
            \frac{(c+\mu A)/\cos(\theta)}{1-{\mu}^{2}\tan^{2}(\theta)}\label{eq:Wlinear}\\
\Delta & = &
T_{instr}(L)+\frac{L\tan(\theta)(1+{\mu}^{2})}{1-{\mu}^{2}\tan^{2}(\theta)}+
            \frac{\mu (c+\mu
            A)\tan(\theta)/\cos(\theta)}{1-{\mu}^{2}\tan^{2}(\theta)}\label{eq:Deltalinear}
\end{eqnarray}

Here we have three unknown ($T_{instr}$, $\mu$ and $c$) and only two equations. We
may exploit the linearity of Eq.~\ref{eq:Wlinear} in the applied load, in the form:
\begin{equation}
W\equiv\mu_{app} L+c_{app} \label{eq:Wlinear_apparent}
\end{equation}
to extract the slope and the offset, both depending on the unknown $\mu$ and $c$.
The effect of a local tilt of the surface is the introduction of an apparent
friction coefficient $\mu_{app}$ and an apparent friction offset $c_{app}$:
\begin{eqnarray}
\mu_{app}(\mu,\theta) & = & \mu\frac{1+\tan^{2}(\theta)}{1-{\mu}^{2}\tan^{2}(\theta)}\label{eq:mu_app}\\
c_{app}(\mu,c,\theta,A) & = & \frac{(c+\mu
A)/cos(\theta)}{1-{\mu}^{2}\tan^{2}(\theta)}\label{eq:c_app}
\end{eqnarray}
The linearity of Eq.~\ref{eq:Wlinear_apparent} is a necessary condition for the
validity of the assumption of the modified Amonton's law (only a necessary, and not
sufficient condition !). Notice that, because of the dependence of $T_{instr}$
on load $L$, we can not exploit Eq.~\ref{eq:Deltalinear} to extract the true values
of $\mu$ and $c$ as for Eq.~\ref{eq:Wlinear}.

In the last equation we have explicitly indicated the dependence of $T_{instr}$ on
the load $L$. If the derivative of $\Delta$ with respect to $L$ is computed, as in the
calibration procedure for the lateral sensitivity $\alpha$ described by Ogletree
{\textit{et al.}}~\cite{ogle96}, the residual term $dT_{instr}/dL$ can make the calibration
procedure inaccurate (see discussion below).

Eq.~\ref{eq:mu_app} represents the well known topographic correction proposed by
Bushan \etal~\cite{sun00} It is a 2nd-order equation whose meaningful root is the
true friction coefficient $\mu$. This topographic correction holds under the
assumptions that the friction is linearly dependent on the load and that the
instrumental offset does not change in Trace and Retrace.

Because of the residual offset due to adhesion, the apparent friction coefficient
$\mu_{app}$ must be calculated via a fit of the $W$ vs. $L$ curve, and not simply
as: $\mu_{app}=W/L$. This offset is present even in case a simple Coulomb-Amonton's
law is assumed (i.e. $c=0$), and follows from considering explicitly the adhesion
force in the decomposition of forces acting on the AFM tip (see
Fig.~\ref{fig:jap3}). In the case of a flat surface ($\theta=0$)
Eq.~\ref{eq:W_without_c} reduces to the Amonton's-Coulomb equation: $f=\mu (L+A)$.

We end this section with some consideration about the application of the equations
of the topographic corrections in the case of adhesive multi-asperity contact for
the calculation of the cantilever lateral sensitivity $\alpha$ proposed by Ogletree \etal~in Ref.~\onlinecite{ogle96}.
Omitting the offset $c$ in Eqs.~\ref{TN_linear}-\ref{eq:Deltalinear}, the following set of equations is obtained:
\begin{eqnarray}
T_{+} & = & L\frac{\mu+\tan(\theta)}{1-\mu\tan(\theta)}+
            \frac{\mu A/L\cos(\theta)}{1-\mu\tan(\theta)}\nonumber\\
N_{+} & = & L\cos(\theta)\frac{1+\tan^{2}(\theta)}{1-\mu\tan(\theta)}
            +\frac{A}{1-\mu\tan(\theta)}\nonumber\\
T_{-} & = & L\frac{\mu-\tan(\theta)}{1+\mu\tan(\theta)}+
            \frac{\mu A/L\cos(\theta)}{1-\mu\tan(\theta)}\nonumber\\
N_{-} & = & L\cos(\theta)\frac{1+\tan^{2}(\theta)}{1+\mu\tan(\theta)}
            +\frac{A}{1+\mu\tan(\theta)}\label{eq:TN_0linear}
\end{eqnarray}
and
\begin{eqnarray}
W & = & L\mu\frac{1+\tan^{2}(\theta)}{1-{\mu}^{2}\tan^{2}(\theta)}+
        \frac{\mu A/\cos(\theta)}{1-\mu^2\tan^2(\theta)}\label{eq:W_without_c}\\
\Delta & = & T_{instr}(L)+
             \frac{L\tan(\theta)(1+{\mu}^{2})}{1-{\mu}^{2}\tan^{2}(\theta)}+
             \frac{\mu^2 A\tan(\theta)/\cos(\theta)}{1-\mu^2\tan^2(\theta)}\label{eq:D_without_c}
\end{eqnarray}

Ogletree \etal ~obtained a set of equations slightly different from our
Eqs.~\ref{eq:W_without_c} and~\ref{eq:D_without_c}, because of the neglecting of
adhesion $A$ in the decomposition of forces along the normal direction. This however
does not affect in principle the calculation of the lateral sensitivity $\alpha$
because this is based on the \textit{derivatives} of Eqs.~\ref{eq:W_without_c}
and~\ref{eq:D_without_c} with respect to $L$. Derivation is supposed to clean off
the offsets, as long as the dependence of $T_{instr}$ on load $L$ in
Eq.~\ref{eq:D_without_c} is weak. More precisely, in Eq.~\ref{eq:D_without_c} the
derivative of $T_{instr}$ versus $L$ must be (much) smaller than the term
$x=tan(\theta)(1+\mu^2)/(1-\mu^2tan^2(\theta))$. The
procedure we have described in section~\ref{section:instr_offset} for the
measurement of the instrumental offset, can be used to inquiry whether the method
proposed by Ogletree \etal ~can be applied. If the derivative of $T_{instr}$ versus
$L$ is comparable to $x$, the laser alignment should be optimized. This procedure
should be followed each time a new cantilever is used, in order to validate the
goodness of the laser alinement and minimize the unwanted effects of cross-talking.
The main drawback of this procedure is that it requires a knowledge of both the tilt
angle of the reference grating and the friction coefficient of the tip-grating pair.

\section{\label{section:protocol_exp}Experimental setup and data analysis}

We have developed a complete quantitative protocol for the quantitative
characterization of the friction properties of corrugated surfaces
using the AFM, based on the equations derived so far.

We have used a Nanoscope Multimode IIIa AFM, that can be housed
in a sealed chamber connected to a humidifier in order to work
in controlled humidity and atmosphere (typically in dry and wet nitrogen). The
deflection setpoint, determining the external applied load, is remotely
controlled by a PC. The application of the load is synchronized with the
end-line and end-frame triggers from the microscope, such that we can acquire a
complete friction vs. load curve in a single AFM scan of typically 512 lines, 512
points per line. The experimental data are exported and post-processed via dedicated
routines. Cantilever force constant $k$ and lateral sensitivity $\alpha$ are
calibrated using the thermal noise method~\cite{butt95,sader98,sta00,levy02} and the
wedge method proposed by Ogletree \etal~\cite{ogle96}. Our calibration
methods do not require any analytical calculation based neither on the
assumption of any oversimplified cantilever geometry nor on SEM measurements, nor on
any knowledge of the elastic properties of the cantilever material.

Our protocol requires the acquisition of four maps: the topographic map, the lateral
force maps acquired in trace and retrace (i.e. the $TR$ and $RETR$ maps in volts)
and the deflection setpoint map proportional to the external applied load
$L$. These four maps should be recorded simultaneously because they are strictly in
one-by-one correspondence. However, only up to two independent signals apart from the
topographic one can be recorded simultaneously by our AFM controller. We record the topographic and lateral
force maps together and the deflection setpoint map separately, because this is a quite
stable and reproducible signal. We scan on the same line during one ramp, and repeat
the ramp on different points. Adhesion is extracted from force versus distance
curves acquired on a flat region of the sample. We repeat the adhesion measurement
many times applying offsets in the \textit{xy} directions and then averaging, in
order to reduce the errors coming from the local corrugation of the
surface. Assuming a constant adhesion across the whole scan window, an adhesion map
$A$ is readily obtained. Tip velocity is typically 1 $\mu m/s$, scan size is up to 1
$\mu m$, and resolution is 512 points per line. In each lateral force map we can
change the applied load up to 512 times. Load range is typically 0-100 nN.

Topographic and lateral force maps are first smoothed using a one-dimensional kernel
whose size equals the tip size (in point units), in order to smooth out all the
information that are below the effective resolution limit. The map of local slopes
$tan(\theta)$ is calculated from the topographic map, with a resolution equal to the
tip radius $R$, so that we have $tan(\theta) \sim tan(\theta)_R$ (see
section~\ref{section:topo_correction}). At this point we calculate $W$ from $TR$ and
$RETR$ (Eq.~\ref{eq:W}). The map of local slopes, the map of external loads $L$, and
the $W$ map are in one-to-one correspondence.
\begin{figure}
 \centering
 \includegraphics[scale=1]{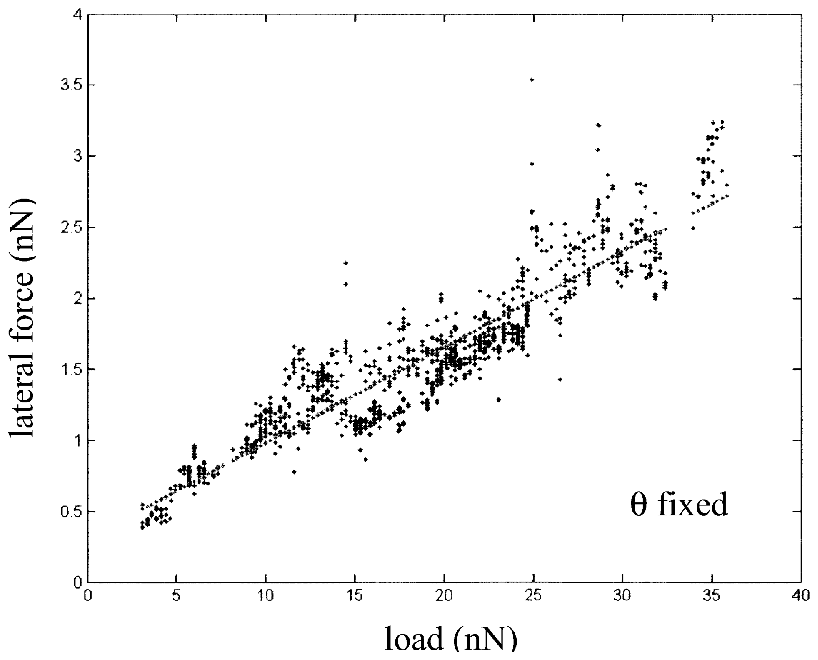}
 \caption{\label{fig:jap5}}
\end{figure}
Making a sampling of the slope map and exploiting the one-to-one correspondence with
other maps, the family of sets $\{ W,L \}_{\theta}$ is obtained, each containing the
pairs $\{ W,L \}$ corresponding to points in the topographic map having the same
slope. Each set gives a $W$ vs. $L$ characteristics (an example is shown in
Fig.~\ref{fig:jap5}) that in turn can be fitted to extract the apparent friction
coefficient $\mu_{app}(\mu,\theta)$ and the apparent adhesive offset
$c_{app}(\mu,c,\theta,A)$.
\begin{figure}[h]
 \centering
 \includegraphics[scale=1]{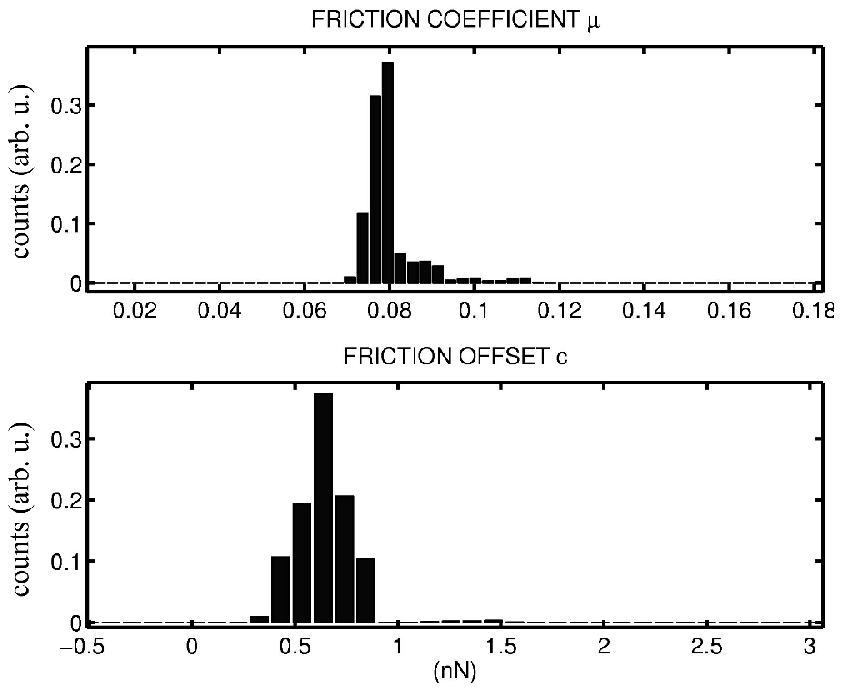}
 \caption{\label{fig:jap6}}
\end{figure}
The linear fit is weighted with errors on both $W$ and $L$ coming from the
indetermination on the vertical force constant $k$, \textit{z}-sensitivity
\textit{z-sens}, and lateral sensitivity $\alpha$ (see
Appendix~\ref{section:error_analysis}). From the apparent friction coefficient
$\mu_{app}$, given the slope $tan(\theta)$, the true friction coefficient $\mu$ is
extracted (Eq.~\ref{eq:mu_app}). From the apparent adhesive offset $c_{app}$, given
the slope, the adhesion map, and the friction coefficient $\mu$, the adhesive offset
$c$ is extracted (Eq.~\ref{eq:c_app}). The propagation of errors is
taken into account in order to determine the error on both $\mu$ and $c$
(Eqs.~\ref{eq:mu_error_final} and~\ref{eq:c_error_final}). The errors calculated for
$\mu$ and $c$ are in turn used to obtain the weighted distributions of measured values
of $\mu$ and $c$ (Fig.~\ref{fig:jap6}). Finally, average values
and standard deviations are obtained fitting the histograms with gaussian curves.

\section{\label{section:tribo_exp}Results and discussion}

In order to illustrate the importance of the topographic correction for the
measurement of the friction coefficient of corrugated surfaces with FFM, we
compare the lateral force dispersion measured by FFM on a Polytetrafluoroethylene
(PTFE) sample with the theoretical predictions of Eq.~\ref{eq:Wlinear}, evaluated using the measured values of the
friction coefficient and adhesive offsets.

\begin{figure}[pt]
\includegraphics[scale=1]{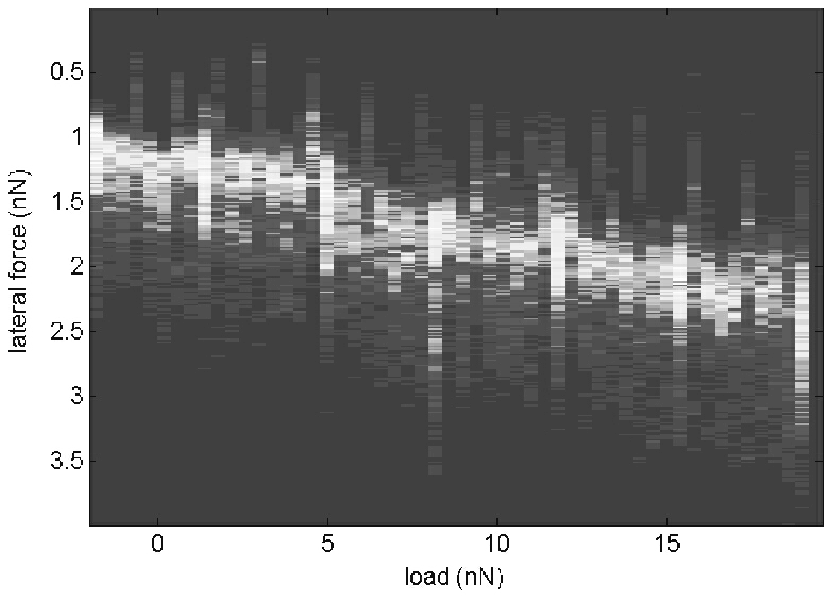}
 \caption{\label{fig:jap7}}
\end{figure}

\begin{figure}[pt]
\includegraphics[scale=1]{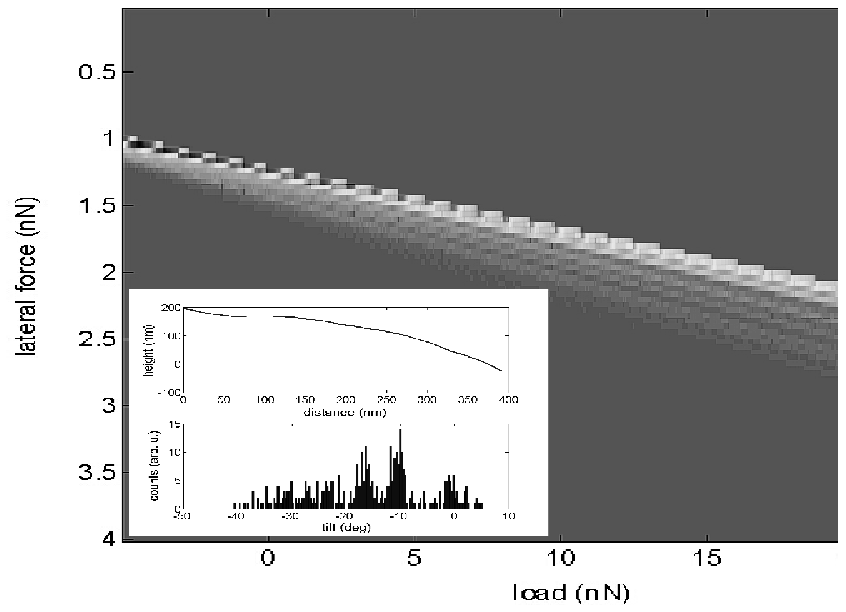}
 \caption{\label{fig:jap8}}
\end{figure}

\begin{figure}[pt]
\includegraphics[scale=1]{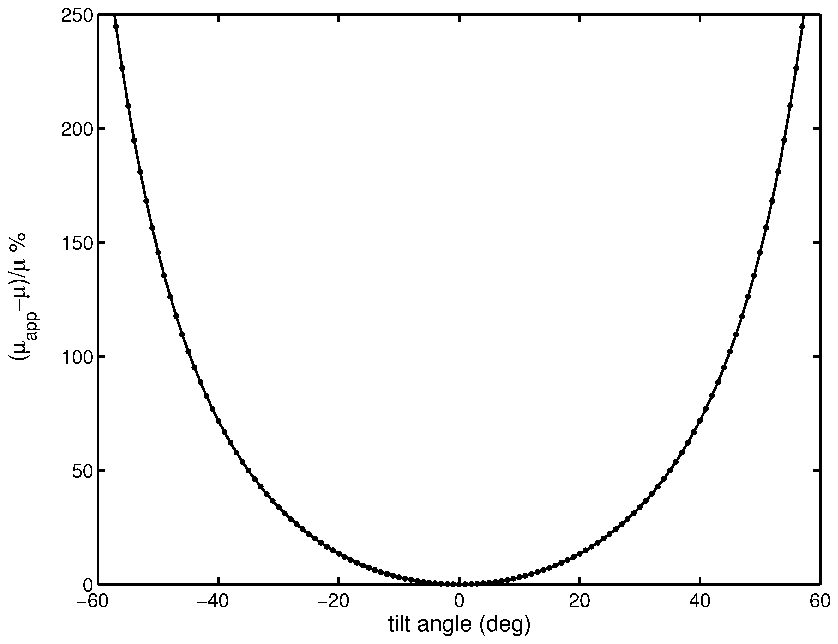}
 \caption{\label{fig:jap9}}
\end{figure}

The dispersion of the measured lateral force values is shown in Fig.~\ref{fig:jap7}.
The friction coefficient and the adhesive offset, as measured with the topographic
correction procedure, are $\mu=0.057 \pm 0.009$ and $c=0.81 \pm 0.17$ nN. Other
parameters are: $\alpha=30$ nN/V, $k=0.1$ N/m, \textit{zsens}=116 nm/V and adhesion
$A=7$ nN. The lateral force dispersion of the PTFE corrugated sample simulated using
Eq.~\ref{eq:Wlinear} is shown in Fig.~\ref{fig:jap8} (the distribution of tilt angles
of the PTFE sample is shown in the inset, together with a topographic profile).
The effect of the presence of a distribution of tilt angles at the sample surface is the
coexistance in Fig.~\ref{fig:jap8} of several linear trends, one stronger (corresponding to
the intense peaks in the tilt angle distribution between -10 and -20 degs) and the others
wicker. In the experimental dispersion we find qualitatively the same trend, although
the statistical noise makes the different linear trends mixing and broadening. The topographic
correction would make all the linear trends collapse into one. Neglecting the
topographic correction would likely lead to inaccurate measurements of the friction
coefficient. In order to quantify the error induced by the presence of a
distribution of tilt angles at the sample surface, we plot in
Fig.~\ref{fig:jap9} the relative error $(\mu_{app}-\mu)/\mu$, calculated from
Eq.~\ref{eq:mu_app}, versus the tilt angle. In the calculation of
$(\mu_{app}-\mu)/\mu$ we assumed $\mu=0.05$, although the
result is largely independent on the value of $\mu$. It can be noticed that for tilt
angles bigger than about 20 degs the discrepancy increases rapidly above 10\% and
becomes larger than 70\% at 40 degs. The slopes in the region of the leading trends
in Fig.~\ref{fig:jap7} span approximately from 0.056 to 0.061.
The difference is of the order of 10\%, which is the expected deviation induced by
tilt angles below 20 degrees. The topographic correction is required as long as the
mean tilt is larger than about 15-20 degrees.

The presence of a distribution of tilt angles amplifies the statistical noise on the
measured values of the lateral force. In addition, it must be considered that often
additional trends are present in the lateral force map, due to those
contributions to friction which are not accounted for by the topographic correction,
such as the ploughing and the collision effect~\cite{sun00}, etc... Even in the case
of small tilt angles, the topographic correction procedure can help in minimizing the
effects of this noise.

The validity of the friction model that we have adopted in the topographic correction
procedure can be verified a posteriori from the linearity of the experimental $W$
vs. $L$ curves and from the values of the adhesive offset $c$. We have found that the
modified Amonton's law is appropriate in describing the frictional behaviour of a
number of systems in ambient conditions (nanostructured carbon films grown via
Supersonic Cluster Beam Deposition~\cite{pod02c}, PTFE-based coatings obtained
through syntherization of nanoemulsions~\cite{pod02b}, and silicon) , for loads
below 100 nN. The measured values of the adhesive offset $c$ were always different
from zero, in the limit of the experimental error. Typical values of the adhesive
offset measured are up to few nanonewtons. If this offset was due to a mere
uncertainty in the measurement of the adhesion force, we should assume a relative
error in its measurement larger than 100\%. For smaller values of $c$, well
below 1 nN, relative errors in $A$ of the order 20-40\% could in fact be compatible with the
simple Coulomb equation (see Eq.~\ref{eq:coulomb}). However, if the adhesion force is
carefully extracted averaging over many force vs. distance curves, we believe that
the error associated can be of the order of 10\%. We are thus confident that our
experimental data support the validity of the modified Amonton's law as an effective
friction model, whose origin must be deeply investigated in the framework of
adhesive contact mechanics. We finally notice that if wear takes place during
scanning, the tip-sample interface and, consequently, the adhesion force would change.
This would represent an additional source of noise, in that the value of
adhesion used in the fit would not correspond to the actual ones. This effect can be
minimized, if not avoided, operating the microscope in the low-load regime.

We point out finally that the relative errors in the measured friction coefficients
and adhesive offsets are typically comparable with that of the factor
$\textit{zsens}\cdot k/\alpha$, which basically convert the friction coefficient
extracted from the experimental data in Volts into the true one. This error is
typically 15-30\%, depending on the accuracy of the cantilever calibration
procedures. This error can be considered systematic and can be reduced only
improving the accuracy of the calibration methods (and this is actually one of the
bottlenecks of nano-friction measurements). The statistical errors in the measured
lateral force values are instead efficiently smoothed out by our protocol, because
of the large number of points that are processed in a single image, the large number
of lateral force maps that can be acquired in a reasonable time.

\section{\label{section:conclusion}Conclusions}

We have discussed the problem of the topographic correction of lateral force maps,
which is necessary to obtain quantitative information on the tribological properties of
corrugated samples via FFM experiments. We have recognized that there are intrinsic
instrumental limitations which reduce the accuracy of these measurements. In
particular, the unreliable measurement of the instrumental lateral force offset
poses severe limitations to the applicability of the topographic correction and
should be taken into account whenever the decomposition of forces on a locally
tilted surface is considered, such as in the procedure for the calibration of the
lateral sensitivity proposed in Ref.~\onlinecite{ogle96}.

Assuming that the friction behavior is well described by a modified Amonton's law,
where a constant offset is explicitly considered in order to account for the
zero-load friction force typical of single asperity adhesive friction, we have
solved the problem of the topographic correction of FFM lateral force maps and
developed a complete protocol, which provide quantitative and reliable
characterization of frictional properties of corrugated materials. The final error
in the measurement is basically due to the uncertainty in the force constants of the
cantilever and it is typically of the order of 15-30\%. Neglecting the topographic
correction on corrugated samples can lead to even larger errors.

The frictional model adopted is appropriate to describe the behavior of
many systems in humid environment. The quantitative assessment of this offset is
important in order to determine the total friction force, for comparison among
different materials.

\appendix*

\section{\label{section:error_analysis}Error analysis}

We extract an apparent friction coefficient $\mu_{app}$ and an apparent offset
$c_{app}$ via a linear fit of the $W$ vs. $L$ curve (Eq.~\ref{eq:Wlinear}). Notice
that the original signals $W$ and $L$, as well as the adhesion force $A$, are given
in Volts. They must be converted in nanoNewton using the cantilever vertical force
constant $k$ in $N/m$ and sensitivity \textit{z-sens} in $nm/V$, and the lateral
sensitivity $\alpha$ in $nN/V$. We assume a relative error of 10\% for $k$ and
$\alpha$ and of 2\% for \textit{z-sens}, which is calculated from the slope of
force-vs.-distance curves taken on a hard flat surface, averaging over many
acquisitions. The relative error for the applied load $L$ is thus 12\%, that for the
lateral force $W$ is 10\%. The linear fit of the $W$ vs. $L$ curve is weighted with
the errors of $W$ and $L$. We use the error of $A$ later in the error propagation
analysis. Notice that the relative error associated to adhesion (15\%) comes from
the force constant (~10\%) and from the statistical error associated to the
extraction of the adhesion force from many force vs. distance curves (~5\%). The
error in \textit{z-sens} does not contribute to the error of $A$ because the depth
of the pull-off region in the approaching-retracting curve is calculated on the
retraction length axis directly in nm with good accuracy. $\mu$ and $c$ are the true
incognitae of the problem, while $A$, $\theta$, $W$ and $L$ are known or measured,
each with its own error. From Eqs.~\ref{eq:mu_app} and~\ref{eq:c_app} we have:
\begin{eqnarray}
\mu(\mu_{app},\theta) & = & \frac{-\frac{1+\tan^{2}(\theta)}{\mu_{app}}+\sqrt{(\frac{1+\tan^{2}(\theta)}{\mu_{app}})^2 + 4\tan^{2}(\theta)}}{2\tan^{2}(\theta)} \label{eq:mu_true}\\
c(\mu,c_{app},\theta,A) & = & c_{app} \cos(\theta)(1-{\mu}^{2}\tan^{2}(\theta)) -\mu
A \label{eq:c_true}
\end{eqnarray}
We apply standard error propagation analysis to Eqs.~\ref{eq:mu_true} and~\ref{eq:c_true} to calculate the errors of $\mu$ and $c$ (the error in $\theta$ is
negligible):
\begin{eqnarray}
\delta\mu^2 & = & (\frac{\delta\mu}{\delta\mu_{app}}) ^2\delta\mu_{app}^2 \label{eq:mu_error}\\
\delta c^2 & = & (\frac{\delta c}{\delta\mu})^2 \delta\mu^2 + (\frac{\delta
c}{\delta c_{app}})^2 \delta c_{app}^2 + (\frac{\delta c}{\delta A})^2 \delta A^2
\label{eq:c_error}
\end{eqnarray}
We calculate first $\frac{\delta\mu}{\delta\mu_{app}}$:
\begin{eqnarray}
\frac{\delta\mu}{\delta\mu_{app}} & = &
                                        \frac{1}{2\tan^2({\theta})}\{\frac{1+\tan^{2}(\theta)}{\mu^2_{app}}
                                        -[(\frac{(1+\tan^{2}(\theta)}{\mu_{app}})^2+4\tan^{2}(\theta)]^{-1/2}
                                        (\frac{(1+\tan^{2}(\theta))^2}{\mu^3_{app}})\}\nonumber\\
                                 & = & \frac{1}{2\tan^2({\theta})}(\frac{1+\tan^{2}(\theta)}{\mu^2_{app}})
                                       \{1-\frac{1+\tan^{2}(\theta)}{\mu_{app}}[(\frac{(1+\tan^{2}(\theta)}{\mu_{app}})^2+4\tan^{2}(\theta))^{-1/2}]\}\nonumber\\
                                 & = & \frac{1}{2\mu_{app}\tan(\theta)}\frac{1+\tan^2(\theta)}{\mu_{app}\tan(\theta)}
                                       \{1-\frac{1+\tan^{2}(\theta)}{\mu_{app}\tan(\theta)}[(\frac{1+\tan^2(\theta)}{\mu_{app}\tan(\theta)})^2+4]^{-1/2}\}\nonumber
\end{eqnarray}
Setting
\begin{equation}
x=\frac{1+\tan^{2}(\theta)}{\mu_{app}\tan(\theta)}
\end{equation}
we obtain:
\begin{equation}
\frac{\delta\mu}{\delta\mu_{app}}=\frac{x}{2\mu_{app}\tan(\theta)}[1-x(x^2+4)^{-1/2}]
\end{equation}
and from Eq.~\ref{eq:mu_error}:
\begin{equation}
\delta\mu^2=(\frac{x}{2\mu_{app}\tan(\theta)})^2[1-x(x^2+4)^{-1/2}]^2
\delta\mu^2_{app} \label{eq:mu_error_final}
\end{equation}
Because $x^2\gg 4$, Eq.~\ref{eq:mu_error_final} can be replaced with:
\begin{equation}
\delta\mu^2=(\frac{1}{1+\tan^2(\theta)})^2 \delta\mu^2_{app}
\label{eq:mu_error_final_short}
\end{equation}
Concerning the error $\delta c^2$, we have:
\begin{eqnarray}
\frac{\delta c}{\delta\mu}      & = & -(A + 2 c_{app}\cos(\theta)\tan^2(\theta) \mu)\\
\frac{\delta c}{\delta c_{app}} & = & \cos(\theta)(1-\mu^2\tan^2(\theta))\\
\frac{\delta c}{\delta A}       & = & -\mu
\end{eqnarray}
and finally:
\begin{equation}
\delta c^2 = \cos^2(\theta)(1-\mu^2\tan^2(\theta))^2  \delta c^2_{app} + (A+2
c_{app}\cos(\theta)\tan^2(\theta) \mu)^2 \delta\mu^2 + \mu^2 \delta A^2
\label{eq:c_error_final}
\end{equation}

\begin{acknowledgments}
The authors would like to thank M. Coletti for its contributions in the development
of the nanotribological protocol.
\end{acknowledgments}


%

\pagebreak

\section*{Figure captions}
\begin{description}
  \item{Figure~\ref{fig:jap1}.} Experimental friction loop. The measured $TR$ and
  $RETR$ signals and their half-sum $\Delta$ and half-difference $W$ are shown.
  \item{Figure~\ref{fig:jap2}.} Schematic representation of the contact
  between the AFM tip and the sample surface, characterized by the
  correlation length $\xi$. The cases $\xi/R\leq1$ (left) and $\xi/R\geq1$ (right)
  are shown. The tilt of the surface calculated with resolution equal to the radius of
  the bigger probe, $\theta^*$, is also shown.
  \item{Figure~\ref{fig:jap3}.} Forces diagram for the AFM tip sliding down a sloped
  topographic feature (retrace).
  \item{Figure~\ref{fig:jap4}.} The instrumental offset $T_{instr}$ as a function of the external applied load.
  \item{Figure~\ref{fig:jap5}.} $\{W,L \}_{\theta}$ characteristics.
  \item{Figure~\ref{fig:jap6}.} Weighted histograms of values of $\mu$ and $c$
  extracted from each $\{W,L\}_{\theta}$ characteristics.
  \item{Figure~\ref{fig:jap7}.} Dispersion of lateral force values measured on the
  PTFE sample without the topographic correction.
  \item{Figure~\ref{fig:jap8}.} Lateral force dispersion of a PTFE sample simulated from Eq.~\ref{eq:Wlinear}
  using topographic and frictional parameters as given by the friction analysis.
  In the inset are shown the topographic profile and the slope distribution of the
  PTFE sample whose lateral force distribution is shown in Fig.~\ref{fig:jap7}.
  \item{Figure~\ref{fig:jap9}.} Relative error in the measured friction coefficient as a function of the local tilt angle $\theta$.
\end{description}

\printfigures

\begin{thebibliography}{42}
\expandafter\ifx\csname natexlab\endcsname\relax\def\natexlab#1{#1}\fi
\expandafter\ifx\csname bibnamefont\endcsname\relax
  \def\bibnamefont#1{#1}\fi
\expandafter\ifx\csname bibfnamefont\endcsname\relax
  \def\bibfnamefont#1{#1}\fi
\expandafter\ifx\csname citenamefont\endcsname\relax
  \def\citenamefont#1{#1}\fi
\expandafter\ifx\csname url\endcsname\relax
  \def\url#1{\texttt{#1}}\fi
\expandafter\ifx\csname urlprefix\endcsname\relax\def\urlprefix{URL }\fi
\providecommand{\bibinfo}[2]{#2}
\providecommand{\eprint}[2][]{\url{#2}}

\bibitem[{mrs(2001)}]{mrsbull01}
\emph{\bibinfo{title}{Materials Research Society Bulletin}},
  vol.~\bibinfo{volume}{26} (\bibinfo{year}{2001}).

\bibitem[{\citenamefont{Bhushan}(1999)}]{bhu99b}
\bibinfo{author}{\bibfnamefont{B.}~\bibnamefont{Bhushan}},
  \emph{\bibinfo{title}{Handbook of {M}icro and {N}ano {T}ribology}}
  (\bibinfo{publisher}{CRC Press}, \bibinfo{year}{1999}).

\bibitem[{\citenamefont{Bhushan}(2001{\natexlab{a}})}]{bhu01}
\bibinfo{author}{\bibfnamefont{B.}~\bibnamefont{Bhushan}},
  \bibinfo{journal}{Proc. Instn. Mech. Engrs} \textbf{\bibinfo{volume}{215}}
  (\bibinfo{year}{2001}{\natexlab{a}}).

\bibitem[{\citenamefont{Carpick and Salmeron}(1997)}]{carp97}
\bibinfo{author}{\bibfnamefont{R.}~\bibnamefont{Carpick}} \bibnamefont{and}
  \bibinfo{author}{\bibfnamefont{M.}~\bibnamefont{Salmeron}},
  \bibinfo{journal}{Chem. Rev.} \textbf{\bibinfo{volume}{97}},
  \bibinfo{pages}{1163} (\bibinfo{year}{1997}).

\bibitem[{\citenamefont{Dedkov}(2000)}]{ded00}
\bibinfo{author}{\bibfnamefont{G.}~\bibnamefont{Dedkov}},
  \bibinfo{journal}{Phys. Stat. Sol. (a)} \textbf{\bibinfo{volume}{179}},
  \bibinfo{pages}{3} (\bibinfo{year}{2000}).

\bibitem[{\citenamefont{Gnecco et~al.}(2001)\citenamefont{Gnecco, Bennewitz,
  Gyalog, and Meyer}}]{gne01}
\bibinfo{author}{\bibfnamefont{E.}~\bibnamefont{Gnecco}},
  \bibinfo{author}{\bibfnamefont{R.}~\bibnamefont{Bennewitz}},
  \bibinfo{author}{\bibfnamefont{T.}~\bibnamefont{Gyalog}}, \bibnamefont{and}
  \bibinfo{author}{\bibfnamefont{E.}~\bibnamefont{Meyer}}, \bibinfo{journal}{J.
  Phys.: Cond. Matt.} \textbf{\bibinfo{volume}{13}}, \bibinfo{pages}{R619}
  (\bibinfo{year}{2001}).

\bibitem[{\citenamefont{Meyer and Amer}(1990)}]{mey90}
\bibinfo{author}{\bibfnamefont{G.}~\bibnamefont{Meyer}} \bibnamefont{and}
  \bibinfo{author}{\bibfnamefont{N.}~\bibnamefont{Amer}},
  \bibinfo{journal}{Appl. Phys. Lett.} \textbf{\bibinfo{volume}{57}},
  \bibinfo{pages}{2089} (\bibinfo{year}{1990}).

\bibitem[{\citenamefont{Grafstrom et~al.}(1993)\citenamefont{Grafstrom,
  Neitzert, Hagen, Ackermann, Neumann, Probst, and Wortge}}]{graf93}
\bibinfo{author}{\bibfnamefont{S.}~\bibnamefont{Grafstrom}},
  \bibinfo{author}{\bibfnamefont{M.}~\bibnamefont{Neitzert}},
  \bibinfo{author}{\bibfnamefont{T.}~\bibnamefont{Hagen}},
  \bibinfo{author}{\bibfnamefont{J.}~\bibnamefont{Ackermann}},
  \bibinfo{author}{\bibfnamefont{R.}~\bibnamefont{Neumann}},
  \bibinfo{author}{\bibfnamefont{O.}~\bibnamefont{Probst}}, \bibnamefont{and}
  \bibinfo{author}{\bibfnamefont{M.}~\bibnamefont{Wortge}},
  \bibinfo{journal}{Nanotechnology} \textbf{\bibinfo{volume}{4}},
  \bibinfo{pages}{143} (\bibinfo{year}{1993}).

\bibitem[{\citenamefont{Labardi
  et~al.}(1994{\natexlab{a}})\citenamefont{Labardi, Allegrini, and
  Fuso}}]{lab94b}
\bibinfo{author}{\bibfnamefont{M.}~\bibnamefont{Labardi}},
  \bibinfo{author}{\bibfnamefont{M.}~\bibnamefont{Allegrini}},
  \bibnamefont{and} \bibinfo{author}{\bibfnamefont{F.}~\bibnamefont{Fuso}},
  \bibinfo{journal}{J. Vac. Sci. Technol.} \textbf{\bibinfo{volume}{B12}},
  \bibinfo{pages}{1642} (\bibinfo{year}{1994}{\natexlab{a}}).

\bibitem[{\citenamefont{Labardi
  et~al.}(1994{\natexlab{b}})\citenamefont{Labardi, Allegrini, Salerno,
  Frediani, and Ascoli}}]{lab94}
\bibinfo{author}{\bibfnamefont{M.}~\bibnamefont{Labardi}},
  \bibinfo{author}{\bibfnamefont{M.}~\bibnamefont{Allegrini}},
  \bibinfo{author}{\bibfnamefont{M.}~\bibnamefont{Salerno}},
  \bibinfo{author}{\bibfnamefont{C.}~\bibnamefont{Frediani}}, \bibnamefont{and}
  \bibinfo{author}{\bibfnamefont{C.}~\bibnamefont{Ascoli}},
  \bibinfo{journal}{Appl. Phys. A} \textbf{\bibinfo{volume}{59}},
  \bibinfo{pages}{3} (\bibinfo{year}{1994}{\natexlab{b}}).

\bibitem[{\citenamefont{Koinkar and Bhushan}(1997)}]{koi97}
\bibinfo{author}{\bibfnamefont{V.}~\bibnamefont{Koinkar}} \bibnamefont{and}
  \bibinfo{author}{\bibfnamefont{B.}~\bibnamefont{Bhushan}},
  \bibinfo{journal}{J. Appl. Phys.} \textbf{\bibinfo{volume}{81}},
  \bibinfo{pages}{2472} (\bibinfo{year}{1997}).

\bibitem[{\citenamefont{Sundararajan and Bushan}(2000)}]{sun00}
\bibinfo{author}{\bibfnamefont{S.}~\bibnamefont{Sundararajan}}
  \bibnamefont{and} \bibinfo{author}{\bibfnamefont{B.}~\bibnamefont{Bushan}},
  \bibinfo{journal}{J. Appl. Phys.} \textbf{\bibinfo{volume}{88}},
  \bibinfo{pages}{4825} (\bibinfo{year}{2000}).

\bibitem[{\citenamefont{Carpick et~al.}(1996)\citenamefont{Carpick, Agrait,
  Ogletree, and Salmeron}}]{car96}
\bibinfo{author}{\bibfnamefont{R.}~\bibnamefont{Carpick}},
  \bibinfo{author}{\bibfnamefont{N.}~\bibnamefont{Agrait}},
  \bibinfo{author}{\bibfnamefont{D.}~\bibnamefont{Ogletree}}, \bibnamefont{and}
  \bibinfo{author}{\bibfnamefont{M.}~\bibnamefont{Salmeron}},
  \bibinfo{journal}{J. Vac. Sci. Tech.} \textbf{\bibinfo{volume}{B14}},
  \bibinfo{pages}{1289} (\bibinfo{year}{1996}).

\bibitem[{\citenamefont{Schwarz et~al.}(1995)\citenamefont{Schwarz, Allers,
  Gensterblum, and Wiesendanger}}]{schw95}
\bibinfo{author}{\bibfnamefont{U.}~\bibnamefont{Schwarz}},
  \bibinfo{author}{\bibfnamefont{W.}~\bibnamefont{Allers}},
  \bibinfo{author}{\bibfnamefont{G.}~\bibnamefont{Gensterblum}},
  \bibnamefont{and}
  \bibinfo{author}{\bibfnamefont{R.}~\bibnamefont{Wiesendanger}},
  \bibinfo{journal}{Phys. Rev. B} \textbf{\bibinfo{volume}{52}},
  \bibinfo{pages}{14976} (\bibinfo{year}{1995}).

\bibitem[{\citenamefont{Enachescu et~al.}(1998)\citenamefont{Enachescu, van~den
  Oetelaar, Carpick, Ogletree, Flipse, and Salmeron}}]{ena98}
\bibinfo{author}{\bibfnamefont{M.}~\bibnamefont{Enachescu}},
  \bibinfo{author}{\bibfnamefont{R.}~\bibnamefont{van~den Oetelaar}},
  \bibinfo{author}{\bibfnamefont{R.}~\bibnamefont{Carpick}},
  \bibinfo{author}{\bibfnamefont{D.}~\bibnamefont{Ogletree}},
  \bibinfo{author}{\bibfnamefont{C.}~\bibnamefont{Flipse}}, \bibnamefont{and}
  \bibinfo{author}{\bibfnamefont{M.}~\bibnamefont{Salmeron}},
  \bibinfo{journal}{Phys. Rev. Lett.} \textbf{\bibinfo{volume}{81}},
  \bibinfo{pages}{1877} (\bibinfo{year}{1998}).

\bibitem[{\citenamefont{Bennewitz et~al.}(2001)\citenamefont{Bennewitz, Gnecco,
  Gyalog, and Meyer}}]{gne01b}
\bibinfo{author}{\bibfnamefont{R.}~\bibnamefont{Bennewitz}},
  \bibinfo{author}{\bibfnamefont{E.}~\bibnamefont{Gnecco}},
  \bibinfo{author}{\bibfnamefont{T.}~\bibnamefont{Gyalog}}, \bibnamefont{and}
  \bibinfo{author}{\bibfnamefont{E.}~\bibnamefont{Meyer}},
  \bibinfo{journal}{Tribology Letters} \textbf{\bibinfo{volume}{10}},
  \bibinfo{pages}{51} (\bibinfo{year}{2001}).

\bibitem[{\citenamefont{Bowden and Tabor}(1950)}]{bow50}
\bibinfo{author}{\bibfnamefont{F.}~\bibnamefont{Bowden}} \bibnamefont{and}
  \bibinfo{author}{\bibfnamefont{D.}~\bibnamefont{Tabor}},
  \emph{\bibinfo{title}{{T}he {F}riction and {L}ubrication of {S}olids}}
  (\bibinfo{publisher}{Clarendon, Oxford}, \bibinfo{year}{1950}).

\bibitem[{\citenamefont{Bhushan}(2001{\natexlab{b}})}]{bhushan01}
\bibinfo{editor}{\bibfnamefont{B.}~\bibnamefont{Bhushan}}, ed.,
  \emph{\bibinfo{title}{{F}undamentals of {T}ribology and {B}ridging the {G}ap
  between {M}acro- and {M}icro/{N}anoscales}}, vol.~\bibinfo{volume}{10} of
  \emph{\bibinfo{series}{{NATO} {ASI} {S}eries}} (\bibinfo{publisher}{Kluwer
  Academic Publishers}, \bibinfo{year}{2001}{\natexlab{b}}).

\bibitem[{\citenamefont{Bhushan}(2001{\natexlab{c}})}]{bhu01b}
\bibinfo{author}{\bibfnamefont{B.}~\bibnamefont{Bhushan}},
  \emph{\bibinfo{title}{{M}odern {T}ribology {H}andbook}},
  vol.~\bibinfo{volume}{1} (\bibinfo{publisher}{CRC Press},
  \bibinfo{year}{2001}{\natexlab{c}}).

\bibitem[{\citenamefont{Barabasi and Stanley}(1995)}]{bar95}
\bibinfo{author}{\bibfnamefont{A.-L.} \bibnamefont{Barabasi}} \bibnamefont{and}
  \bibinfo{author}{\bibfnamefont{H.~E.} \bibnamefont{Stanley}},
  \emph{\bibinfo{title}{Fractal Concepts in Surface Growth}}
  (\bibinfo{publisher}{University Press}, \bibinfo{address}{Cambridge},
  \bibinfo{year}{1995}).

\bibitem[{\citenamefont{Adams and Nosonovsky}(2000)}]{ada00}
\bibinfo{author}{\bibfnamefont{G.}~\bibnamefont{Adams}} \bibnamefont{and}
  \bibinfo{author}{\bibfnamefont{M.}~\bibnamefont{Nosonovsky}},
  \bibinfo{journal}{J. Trib.} \textbf{\bibinfo{volume}{33}},
  \bibinfo{pages}{431} (\bibinfo{year}{2000}).

\bibitem[{\citenamefont{Schwarz et~al.}(1997)\citenamefont{Schwarz, Zworner,
  Koster, and Wiesendanger}}]{schw97}
\bibinfo{author}{\bibfnamefont{U.}~\bibnamefont{Schwarz}},
  \bibinfo{author}{\bibfnamefont{O.}~\bibnamefont{Zworner}},
  \bibinfo{author}{\bibfnamefont{P.}~\bibnamefont{Koster}}, \bibnamefont{and}
  \bibinfo{author}{\bibfnamefont{R.}~\bibnamefont{Wiesendanger}},
  \bibinfo{journal}{Phys. Rev. B} \textbf{\bibinfo{volume}{56}},
  \bibinfo{pages}{6987} (\bibinfo{year}{1997}).

\bibitem[{\citenamefont{Ruan and Bhushan}(1994)}]{rua94}
\bibinfo{author}{\bibfnamefont{J.-A.} \bibnamefont{Ruan}} \bibnamefont{and}
  \bibinfo{author}{\bibfnamefont{B.}~\bibnamefont{Bhushan}},
  \bibinfo{journal}{J. Trib.} \textbf{\bibinfo{volume}{116}},
  \bibinfo{pages}{378} (\bibinfo{year}{1994}).

\bibitem[{\citenamefont{Ogletree et~al.}(1996)\citenamefont{Ogletree, Carpick,
  and Salmeron}}]{ogle96}
\bibinfo{author}{\bibfnamefont{D.}~\bibnamefont{Ogletree}},
  \bibinfo{author}{\bibfnamefont{R.}~\bibnamefont{Carpick}}, \bibnamefont{and}
  \bibinfo{author}{\bibfnamefont{M.}~\bibnamefont{Salmeron}},
  \bibinfo{journal}{Rev. Sci. Instr.} \textbf{\bibinfo{volume}{67}},
  \bibinfo{pages}{3298} (\bibinfo{year}{1996}).

\bibitem[{\citenamefont{Israelachvili}(1992)}]{isr92}
\bibinfo{author}{\bibfnamefont{J.}~\bibnamefont{Israelachvili}},
  \emph{\bibinfo{title}{Intermolecular and surface forces}}
  (\bibinfo{publisher}{Academic Press}, \bibinfo{year}{1992}).

\bibitem[{\citenamefont{Lantz et~al.}(1997)\citenamefont{Lantz, O'Shea, and
  Welland}}]{lan97}
\bibinfo{author}{\bibfnamefont{M.}~\bibnamefont{Lantz}},
  \bibinfo{author}{\bibfnamefont{S.}~\bibnamefont{O'Shea}}, \bibnamefont{and}
  \bibinfo{author}{\bibfnamefont{M.}~\bibnamefont{Welland}},
  \bibinfo{journal}{Phys. Rev. B} \textbf{\bibinfo{volume}{56}},
  \bibinfo{pages}{15345} (\bibinfo{year}{1997}).

\bibitem[{\citenamefont{Frantz et~al.}(1997)\citenamefont{Frantz,
  Artsyukhovich, Carpick, and Salmeron}}]{fra97}
\bibinfo{author}{\bibfnamefont{P.}~\bibnamefont{Frantz}},
  \bibinfo{author}{\bibfnamefont{A.}~\bibnamefont{Artsyukhovich}},
  \bibinfo{author}{\bibfnamefont{R.}~\bibnamefont{Carpick}}, \bibnamefont{and}
  \bibinfo{author}{\bibfnamefont{M.}~\bibnamefont{Salmeron}},
  \bibinfo{journal}{Langmuir} \textbf{\bibinfo{volume}{13}},
  \bibinfo{pages}{5957} (\bibinfo{year}{1997}).

\bibitem[{\citenamefont{Putman and Igarashi}(1995)}]{put95}
\bibinfo{author}{\bibfnamefont{C.}~\bibnamefont{Putman}} \bibnamefont{and}
  \bibinfo{author}{\bibfnamefont{M.}~\bibnamefont{Igarashi}},
  \bibinfo{journal}{Appl. Phys. Lett.} \textbf{\bibinfo{volume}{66}},
  \bibinfo{pages}{1} (\bibinfo{year}{1995}).

\bibitem[{\citenamefont{Fuller and Tabor}(1975)}]{ful75}
\bibinfo{author}{\bibfnamefont{K.}~\bibnamefont{Fuller}} \bibnamefont{and}
  \bibinfo{author}{\bibfnamefont{D.}~\bibnamefont{Tabor}},
  \bibinfo{journal}{Proc. R. Soc. Lond. A} \textbf{\bibinfo{volume}{345}},
  \bibinfo{pages}{327} (\bibinfo{year}{1975}).

\bibitem[{\citenamefont{Bush et~al.}(1976)\citenamefont{Bush, Gibson, and
  Keogh}}]{bus76}
\bibinfo{author}{\bibfnamefont{A.}~\bibnamefont{Bush}},
  \bibinfo{author}{\bibfnamefont{R.}~\bibnamefont{Gibson}}, \bibnamefont{and}
  \bibinfo{author}{\bibfnamefont{G.}~\bibnamefont{Keogh}},
  \bibinfo{journal}{Mechanics Research Communications}
  \textbf{\bibinfo{volume}{3}}, \bibinfo{pages}{169} (\bibinfo{year}{1976}).

\bibitem[{\citenamefont{Kendall}(1986)}]{ken86}
\bibinfo{author}{\bibfnamefont{K.}~\bibnamefont{Kendall}},
  \bibinfo{journal}{Science} \textbf{\bibinfo{volume}{319}},
  \bibinfo{pages}{203} (\bibinfo{year}{1986}).

\bibitem[{\citenamefont{Persson and Tosatti}(2001)}]{per01}
\bibinfo{author}{\bibfnamefont{B.}~\bibnamefont{Persson}} \bibnamefont{and}
  \bibinfo{author}{\bibfnamefont{E.}~\bibnamefont{Tosatti}},
  \bibinfo{journal}{Journal of Chemical Physics} \textbf{\bibinfo{volume}{115}}
  (\bibinfo{year}{2001}).

\bibitem[{\citenamefont{Chow}(2001)}]{cho01}
\bibinfo{author}{\bibfnamefont{T.}~\bibnamefont{Chow}}, \bibinfo{journal}{Phys.
  Rev. Lett.} \textbf{\bibinfo{volume}{86}}, \bibinfo{pages}{4592}
  (\bibinfo{year}{2001}).

\bibitem[{\citenamefont{Johnson et~al.}(1971)\citenamefont{Johnson, Kendall,
  and Roberts}}]{joh71}
\bibinfo{author}{\bibfnamefont{K.}~\bibnamefont{Johnson}},
  \bibinfo{author}{\bibfnamefont{K.}~\bibnamefont{Kendall}}, \bibnamefont{and}
  \bibinfo{author}{\bibfnamefont{A.}~\bibnamefont{Roberts}},
  \bibinfo{journal}{Proc. R. Soc. Lond.} \textbf{\bibinfo{volume}{A324}},
  \bibinfo{pages}{301} (\bibinfo{year}{1971}).

\bibitem[{\citenamefont{Derjaguin et~al.}(1971)\citenamefont{Derjaguin, Muller,
  and Toporov}}]{der71}
\bibinfo{author}{\bibfnamefont{B.}~\bibnamefont{Derjaguin}},
  \bibinfo{author}{\bibfnamefont{V.}~\bibnamefont{Muller}}, \bibnamefont{and}
  \bibinfo{author}{\bibfnamefont{Y.~P.} \bibnamefont{Toporov}},
  \bibinfo{journal}{J. Coll. Int. Sci.} \textbf{\bibinfo{volume}{53}},
  \bibinfo{pages}{314} (\bibinfo{year}{1971}).

\bibitem[{\citenamefont{Schwarz et~al.}(1996)\citenamefont{Schwarz, Bluhm,
  Holscher, Allers, and Wiesendanger}}]{schw96}
\bibinfo{author}{\bibfnamefont{U.}~\bibnamefont{Schwarz}},
  \bibinfo{author}{\bibfnamefont{H.}~\bibnamefont{Bluhm}},
  \bibinfo{author}{\bibfnamefont{H.}~\bibnamefont{Holscher}},
  \bibinfo{author}{\bibfnamefont{W.}~\bibnamefont{Allers}}, \bibnamefont{and}
  \bibinfo{author}{\bibfnamefont{R.}~\bibnamefont{Wiesendanger}}, in
  \emph{\bibinfo{booktitle}{Physics of sliding friction}}, edited by
  \bibinfo{editor}{\bibfnamefont{B.}~\bibnamefont{Persson}} \bibnamefont{and}
  \bibinfo{editor}{\bibfnamefont{E.}~\bibnamefont{Tosatti}}
  (\bibinfo{publisher}{Kluwer Academic Publishers, Netherlands},
  \bibinfo{year}{1996}), p. \bibinfo{pages}{369}.

\bibitem[{\citenamefont{Butt and Jaschke}(1995)}]{butt95}
\bibinfo{author}{\bibfnamefont{H.-J.} \bibnamefont{Butt}} \bibnamefont{and}
  \bibinfo{author}{\bibfnamefont{M.}~\bibnamefont{Jaschke}},
  \bibinfo{journal}{Nanotechnology} \textbf{\bibinfo{volume}{6}},
  \bibinfo{pages}{1} (\bibinfo{year}{1995}).

\bibitem[{\citenamefont{Sader}(1998)}]{sader98}
\bibinfo{author}{\bibfnamefont{J.}~\bibnamefont{Sader}}, \bibinfo{journal}{J.
  Appl. Phys.} \textbf{\bibinfo{volume}{84}}, \bibinfo{pages}{64}
  (\bibinfo{year}{1998}).

\bibitem[{\citenamefont{Stark et~al.}(2000)\citenamefont{Stark, Drobek, and
  Heckl}}]{sta00}
\bibinfo{author}{\bibfnamefont{R.}~\bibnamefont{Stark}},
  \bibinfo{author}{\bibfnamefont{T.}~\bibnamefont{Drobek}}, \bibnamefont{and}
  \bibinfo{author}{\bibfnamefont{W.}~\bibnamefont{Heckl}},
  \bibinfo{journal}{Ultramicroscopy} \textbf{\bibinfo{volume}{86}},
  \bibinfo{pages}{207} (\bibinfo{year}{2000}).

\bibitem[{\citenamefont{Levy and Maaloum}(2002)}]{levy02}
\bibinfo{author}{\bibfnamefont{R.}~\bibnamefont{Levy}} \bibnamefont{and}
  \bibinfo{author}{\bibfnamefont{M.}~\bibnamefont{Maaloum}},
  \bibinfo{journal}{Nanotechnology} \textbf{\bibinfo{volume}{13}},
  \bibinfo{pages}{33} (\bibinfo{year}{2002}).

\bibitem[{\citenamefont{Podest{\`a}
  et~al.}(2002{\natexlab{a}})\citenamefont{Podest{\`a}, Fantoni, Milani,
  Ragazzi, Donadio, and Colombo}}]{pod02c}
\bibinfo{author}{\bibfnamefont{A.}~\bibnamefont{Podest{\`a}}},
  \bibinfo{author}{\bibfnamefont{G.}~\bibnamefont{Fantoni}},
  \bibinfo{author}{\bibfnamefont{P.}~\bibnamefont{Milani}},
  \bibinfo{author}{\bibfnamefont{M.}~\bibnamefont{Ragazzi}},
  \bibinfo{author}{\bibfnamefont{D.}~\bibnamefont{Donadio}}, \bibnamefont{and}
  \bibinfo{author}{\bibfnamefont{L.}~\bibnamefont{Colombo}},
  \bibinfo{journal}{J. Nanosci. Nanotechnol.}
  (\bibinfo{year}{2002}{\natexlab{a}}).

\bibitem[{\citenamefont{Podest{\`a}
  et~al.}(2002{\natexlab{b}})\citenamefont{Podest{\`a}, Fantoni, Milani, Guida,
  and Volponi}}]{pod02b}
\bibinfo{author}{\bibfnamefont{A.}~\bibnamefont{Podest{\`a}}},
  \bibinfo{author}{\bibfnamefont{G.}~\bibnamefont{Fantoni}},
  \bibinfo{author}{\bibfnamefont{P.}~\bibnamefont{Milani}},
  \bibinfo{author}{\bibfnamefont{C.}~\bibnamefont{Guida}}, \bibnamefont{and}
  \bibinfo{author}{\bibfnamefont{S.}~\bibnamefont{Volponi}},
  \bibinfo{journal}{{T}hin {S}olid {F}ilms} \textbf{\bibinfo{volume}{419}},
  \bibinfo{pages}{154} (\bibinfo{year}{2002}{\natexlab{b}}).

\end{thebibliography}
\end{document}